# Towards an Intelligent Framework for Pressure-based 3D Curve Drawing

Chan-Yet Lai and Nordin Zakaria

Universiti Teknologi PETRONAS

**ABSTRACT**

Pen pressure is an input channel typically available in tablet pen device. To date, little attention has been paid to the use of pressure in the domain of graphical interaction, its usage largely limited to drawing and painting program, typically for varying brush characteristic such as stroke width, opacity and color. In this paper, we explore the use of pressure in 3D curve drawing. The act of controlling pressure using pen, pencil and brush in real life appears effortless, but to mimic this natural ability to control pressure using a pressure sensitive pen in the realm of electronic medium is difficult. Previous pressure based interaction work have proposed various signal processing techniques to improve the accuracy in pressure control, but a one-for-all signal processing solution tend not to work for different curve types. We propose instead a framework which applies signal processing techniques tuned to individual curve type. A neural network classifier is used as a curve classifier. Based on the classification, a custom combination of signal processing techniques is then applied. Results obtained point to the feasibility and advantage of the approach.

**Index Terms**: H.5.2[User Interfaces]: Interaction styles; I.3.6[Methodology and Techniques]: Interaction techniques

## 1 INTRODUCTION

Specifying non planar 3D curves from digitized free-form 2D sketched stroke is of fundamental importance in free-form modeling and sketch-based modeling. Although much work has been done in 2D curve sketching, existing methods on 3D curve sketching [1-5] are limited to a plane, indirect and require user to draw from multiple camera positions. In addition, some of the existing approaches require expensive and space-consuming hardware such as magnetic tracking system, stereo glasses, digital projector, and table like rear projection system. Thus it is essential to seek for a simpler 3D curve sketching technique that requires only less costly hardware.

The hardware we focus on for the study is the pen device commonly used in tablet system. Of interest especially within the context of this paper is its ability to provide pressure input. To the best of our knowledge, there has been little work on the use of pressure as an additional input channel in 3D curve drawing. Commonly, most of the previous studies [6-10] explored the use of pressure to perform discrete target selection task. A few studies [11-13] explored the use of pressure control in concert with mouse/pen movement for high precision parameter control [11], concurrent execution of selection and an action [12], and simultaneous manipulation of object orientation and translation [13].

Little work has been done on using pressure for 3D curve drawing as it is difficult to control the pressure through a pressure sensitive pen. Previous work have proposed various signal processing techniques to improve the accuracy in pressure control, but those one-for-all signal processing solutions fail to support different types of curve. There is a need for fit-for-purpose signal processing techniques for each type of sketched curve.

In this paper, we explore the use of pen pressure in concert with pen movement in x-y space to generate 3D curve from a single viewpoint. We assume a user interaction technique where user sketches a 2D free-form stroke with varying pen pressure to specify 3D curve. The pen movement in x-y space depicts the graphical expression of the curve in x-y space, while the pen pressure is used to control the depth profile of the curve in z-space. The amount of pressure applied during sketching is indicated by continuous real-time feedback that mimics the effect of a paint brush painting the water color.

We propose a framework which incorporates pressure based curve drawing technique that is capable of processing the recognized curve with its best combination of signal processing techniques to deal with the unique pressure interaction issues for each type of curve. The key to this capability is a neural network classifier that has been trained to classify curves based on its pressure profile, allowing for processing steps to be applied that has been tuned for each curve class. Due to the use of neural network, an 'artificial intelligence' device, in a sense, the framework we propose here is 'intelligent'.

## 2 RELATED WORK

### 2.1 Curve Drawing

Drawing 3D curve is crucial during the creation and manipulation of surfaces. Consequently 3D curve drawing is an elemental operation that requires extensive research. [3]. Current sketch-based modeling user interfaces rely heavily on 2D and 3D curve specification to specify skeletal shape for implicit surfaces, define control for object deformation, and specify motion camera path. Traditionally, 3D curve specification is performed through controlled points. It is a tedious procedure and required certain mathematical knowledge to obtain the desired complex geometric surface [3]. Many authors regard sketched curve as one of the vital features in modeling tools. Though it is imprecise, it allows fast specification of 3D curves. Therefore 3D curve creation from digitized sketched 2D pen stroke is the elementary aspect of any sketch-based modeling system [1-3, 5]. Previous works on 3D curve specification have widely explored the use of 2D input devices to specify 3D curves. However, the 3D curve specification techniques proposed in previous work are mostly expensive (due to the hardware required), or limited to a plane and needing multiple viewpoints or multiple lines. By far Bae et al. [1] *ILoveSketch* is the only holistic system that allows the sketch of complicated 3D curve networks in a sketch-like workflow. Nevertheless the system is difficult to learn and it is meant for professional product designer. A simple 3D curve specification method was proposed by Cohen et al. [2] where 3D curve is specified by drawing a 2D curve from one viewpoint with its corresponding shadow on the floor plane. This simple technique is intuitive, however specifying the shadow of the 3D curve can be difficult when the 3D curve and 3D scene get complicated. In another research work by De Amicis et al. [3], direct 3D curve specification is achieved through a pencil and rubber metaphor within a semi-immersive environment known as *Virtual Table*. Grossman et al. [4] presented a system that mimics the tape drawing interaction technique from the automobile industry to create digitized non planar 3D curves through a series of 2D curves. Both [3] and [4] require special devices and complex set up.

## 2.2  Pressure Based Interaction

To design an interaction technique that is sensitive to the force applied through a pen, we need to consider the type of pressure sensor used and the user's ability to comfortably control the pressure values. Raw pressure data comes in the form of large number of discrete values. As force is applied to a pressure sensitive stylus, analog force data is produced. This analog force data is then converted into large number of discrete digital values through the Analog to Digital Converter (A-to-D) [6, 10]. Empirical data on user's ability to control pressure has shown that human is not capable of differentiating the granularity of this range of pressure values. This is particularly obvious at both low and high pressure spectrum. The pressure based interaction studies done by researchers since 1996 [6-9, 14] have all reported that at low pressure spectrum, the difference in pressure levels perceived is felt to be too subtle but the difference in pressure level sensed by the digital instrument is far greater than that the user would have expected. At high pressure spectrum, finger tips tremor has the tendency to exert unintended minor force that is not noticeable by human. This unnoticeable minor force variation at high pressure spectrum produces magnified variation in pressure signal. Pressure signal is noisy and it can significantly reduce the accuracy of pressure interaction. Ramos [15] reported that the main sources of noise are input device background noise and physical environment noise, and the noise generated from the unnoticeable nature of hand tremor motion while handling the pressure sensitive pen. Ramos & Balakrishnan [11] proposed the use of low pass filter and hysteresis filter to mitigate signal noise. This straightforward signal filtering scheme work very well in stabilizing the pressure signal in [11]. Nonetheless they concluded that there is still room for improvement by using more sophisticated filtering techniques. Common strategies to improve accuracy in pressure control include the use of a combination of noise filtering techniques to stabilize the pressure signal noise [11], the experiment of various transfer function to discretize the raw pressure values [6, 9-11, 13], the experimental studies to identify the maximum number of discrete pressure level [6-8, 10], the design of algorithm to identify as accurately as possible the exact movement user performs the lifting action to estimate the last intended pressure level [11], the study and design of visual feedback to provide continuous indicator of how much pressure is applied [7, 8, 11, 13, 15], and last but not least, the study of human's perception of pressure [7, 8, 11, 14, 15]. In summary, existing work on pressure-based interaction [6-11, 13-15] concluded that the adequate control of pressure is tightly coupled to the choice of pressure signal stabilization techniques, transfer function, the maximum number of discrete pressure level, and visual feedback.

## 2.3  Classification of Waveform Using Neural Network

An artificial neural network, modeled after the biological brain, has for the past decades enjoyed very extensive applications in various domains that require automated recognition or classification. Learning is accomplished through training algorithms developed based on learning rules presumed to mimic the learning mechanisms of biological systems [16]. The knowledge gained from learning experience is simply the optimum set of connection weights for the particular input patterns presented to the neural network during training. The obtained connection weights will be used to classify fresh input data that the neural network (NN) has never learnt before. The decision making in NN is holistic based on cumulative input patterns [17], thus giving it the ability to generalize. This means a properly trained NN is able to correctly classify data which is out of the training dataset. Due to its unique capability to generalize in the presence of noise, NN has been used widely in signal or waveform analysis in the field of medical [16-25] and finance [26]. Neural network has been explored in the financial market for Elliot waveform recognition [26] to identify and predict repeating pattern in future trends. The application of NN in the medical field is typically for bio signal or waveform pattern recognition and classification to diagnose disease. These include electrocardiogram (ECG) waveform, Doppler signals and electroencephalograph (EEG) signals. These signals and waveform are a record of the propagation of electrical potential generated by different parts of human body cells. They are representative signals that contain valuable information to the nature of the disease, which is reflected in the shape of the waveform [16, 17, 20-23, 25]. The common characteristics of these bio-signals are that they are non stationary, contaminated with noise, and have large variation in the morphologies of the waveform not only of different patients or patient groups but also within the same patient [17, 21, 22, 25]. Numerous other waveform classifier such as beat classifier, digital filter, linear and non linear methods have been explored previously, all perform well on training data but generalize poorly [21, 22, 24, 25]. The unique ability of neural network to generalize the variation of waveform morphology has made neural network a preferred and reliable waveform classifier. A great variety of neural networks have been experimented in previous work for effective classification of ECG [17, 20-25], EEG [16], Doppler [18, 19], and Elliot waveform [26]. All reported good recognition rate with correct classification in over 90% of the cases. On the other hand, classification rate as high as 100% was reported in previous work on Doppler signal classification [18, 19] and Elliot waveform recognition [26]. Similar to the bio signal described earlier, sketched curve pressure profile is noisy and there is a large variation in the pressure profile, not only of different sketch patterns but also within the same sketch pattern. Successful application of neural network in previous works [16-26] for bio signal and waveform classification with correct classification rate in over 90% of the cases and some as high as 100% have demonstrated that the generalization ability of neural network is relatively robust in the presence of noise and the variation of waveform.

## 3  PRESSURE BASED 3D CURVE SKETCHING

We assume the 3D curve interaction technique in [27].The technique uses pen pressure data to directly specify a 3D curve. Users make use of pen pressure to control the curvature (depth profile) of a 3D curve by drawing a 2D curve with varying thickness. The amount of pressure applied during sketching is indicated by the thickness of the 2D curve. When more pressure is applied the 2D curve is thicker and vice versa. The system maps the thickness at each point along the 2D curve to a depth distance. Hence, the thicker the 2D curve at a particular point, the closer the corresponding 3D point to the camera. The generated 3D curve can be further refined by increasing or decreasing the 2D curve thickness, hence closer or further depth distance. With this technique an approximate 3D curve can be sketched and edited directly without much learning is needed. More details on the interaction scheme is in [27].

## 4 Pressure based 3D curve recognition using neural network

Existing pressure based interaction studies have established various signal processing techniques to improve pressure control. We build upon the existing work by incorporating a neural network recognizer that channels curves to the various signal processing techniques proposed in previous pressure based interaction studies. The pressure based sketched curve is recognized through their pressure profiles. Once a pressure based sketched curve is recognized or classified, the 3D curve is refined with the combination of pressure signal processing techniques designated for that particular curve class.

Through an experimental study, we observed that a particular signal processing technique or a combination of signal processing techniques works well to improve pressure control for a particular curve type. Therefore the type of signal processing technique to be applied is dependent on the sketched curve. Sketched curve is freeform strokes; naturally there is a large variation in the pressure based sketched curve pressure profile even for the same sketch pattern. For the purpose of this study, the pressure based sketched curves are classified into three categories; namely spiral curve, forward curve, and backward curve. We limit the spiral to only having one circular shape. A forward curve is a 3D stroke with curve bending towards the camera and hence towards the user. While a backward curve is a 3D stroke with curve bending away from the camera or away from the user. The choice of curve category is inspired by the naturally formed curves commonly found in botanical shape, such as flowers and trees.

### 4.1 Training Data

The pressure signals acquisition was conducted in our research lab. A total of 5 volunteers comprise of 3 males and 2 females ranging in age from 20 to 33 years old were requested to sketch multiple curves for each curve category. The volunteers are of computer science background and only one of them has experience with the tablet used in the experiment, the rest had little or no prior experience. All are right-handed. Prior to the trial, participants were briefed and demonstrated on how to make use of the pen pressure to produce spiral, forward, and backward curve. Participants were given ample time to test and practice; to allow them to perform the required tasks confidently and comfortably. A total of 181 pressure profiles were recorded, which consist of 49 spiral curves, 65 forward curves, and 67 backward curves. All 181 pressure profiles were used as training data for the neural network. Though the pressure profiles within each curve category share some similarities, large variation in the shape of the waveform was observed not only among different individuals but within the same individual as well. We also observed that the number of pressure signal for curves within the same curve category can be ranging from as few as 300 pressure signals to as many as 1400 pressure signals. This phenomenon is attributed to the imprecise free-form sketched curve, where some curves were drawn shorter (smaller) and some were drawn longer (bigger). Fast Fourier Transform (FFT) was used for pre-conditioning the pressure signals. Only the real component of the FFT value was extracted to be used as input data.

### 4.2 Neural Network Experiment

The most popular approach to find the optimum number of hidden layers and hidden neurons is via trial and error. By far, this is the only method used by previous studies on waveform pattern recognition [16-19, 21-23] to find the optimum network topology. In our study, the optimum number of hidden layers and hidden neurons are obtained via experiment. The architecture of the neural network was examined using one and two hidden layers with variable number of hidden neurons. The experiment started with one hidden layer and one hidden neuron, subsequently the number of hidden neurons is increased to observe the error rate. The number of hidden neuron experimented for one hidden layer is ranging from 1 to 100, with an increment of 1 neuron in each training cycle. In two hidden layer experiments, the number of hidden neuron experimented in both hidden layers is ranging from 2 to 100 neurons, with an increment of 2 neurons in one of the hidden layer in each subsequent training cycle, while keeping the number of neuron in the other hidden layer constant. The stopping criteria used for the experiments are 30000 iterations and target error rate of 0.0001

The input layer of the network consists of 50 input neurons to take in the 50 normalized FFT values and the output layer has 3 neurons, each represent a curve class. The network is trained to identify 3 classes of curve patterns. The output value of each output neuron is either logic 1 or logic 0. The binary output set for each curve class is decoded as 001 for backward curve, 010 for forward curve, and 100 for spiral curve. Most of the proposed neural network structure in previous work on waveform pattern recognition [16, 19, 22, 24-26] was trained by backpropagation algorithm. Our neural network was trained using adaptive batch training algorithm. From the preliminary experiment using our training data set, this training algorithm produced lower error rate than the conventional backpropagation training algorithm.

The architecture that produces the lowest error rate from the experiments performed on one hidden layer is 88 neurons. We also observed that even with less number of neurons, as few as 35 neurons, the network is capable of achieving acceptable low error rate. Numerous articles on neural network proposed that the optimum number of neurons for a network would be 2/3 of the summation of input neurons and output neurons, i.e. 2/3*(total input neurons + total output neurons). In our case, this comes to a figure of 35, which is exactly the lowest number of neuron that gave the lowest possible acceptable error rate. From the two hidden layer experimental results obtained, the variation of error rate shows great reduction. The network configuration that produces the lowest error rate is 62 neurons in hidden layer 1 and 46 neurons in hidden layer 2. The best three neural network architectures identified through experiments are i) 50:35:3, ii) 50:88:3 and iii) 50:62:46:3. Few more thousands of trainings were performed using these three configurations until the best possible network architecture and connections weights with the lowest error rates are obtained. The lowest error rates achieved by these three architectures are 0.0100, 0.0088748243, and 0.0002437827 respectively, as shown in Table1.

## 4.3 Pressure Based 3D Sketched Curve Recognition Test

We performed curve recognition tests using the weight connections of the best three neural network architectures. A total of 300 sketched curves were tested; 100 backward curves, 100 forward curves and 100 spiral curves. The same 300 sketched curves were used in the tests performed on all three architectures. Table 1 compares the correct classification rate of the three architectures and their corresponding training error rate. All three architectures share the same classification trend, with backward curve recognized the most and spiral curve recognized the least. We observed that even though the architecture with 35 neurons in one hidden layer has the highest training error rate, surprisingly it is able to recognize the curves better than the other two architectures which have lower training error rates. It can be seen from Table 1, one hidden layer and less neuron architecture is able to achieve higher classification rate than architectures with more neuron and hidden layer.

Table1: Comparison of training error rate and classification result

| Optimum Architecture | Training Error Rate | Classification Accuracy | Test |
|---|---|---|---|
| 50:35:3 | 0.010 | backward curve | 100% |
|  |  | forward curve | 98% |
|  |  | spiral curve | 95% |
| 50:88:3 | 0.00887 | backward curve | 100% |
|  |  | forward curve | 98% |
|  |  | spiral curve | 84% |
| 50:62:46:3 | 0.00024 | backward curve | 100% |
|  |  | forward curve | 86% |
|  |  | spiral curve | 83% |

Neural network architecture with higher number of hidden neuron and hidden layer requires more computations. In our application, real time feedback is essential. When the level of classification accuracy is comparable, the choice of the optimum neural network architecture will be determined by the level of computation complexity. Therefore, we have selected 50:35:3 architecture as the optimum neural network architecture. From the test result we conclude that the one hidden layer with 35 neurons architecture is able to correctly classify 95% and above of the pressure based sketched curve. The detail classification result is presented in Table 2. We are able to achieve 100% correct classification rate for backward curve. As high as 98% of the forward curve is recognized correctly. Spiral curves are classified successfully with 95% correct classification rate. These result show that neural network is able to classify pressure based sketched curve successfully.

From the classification result trend, we observed that backward curve is the most recognize curve, follow by forward curve. Spiral curve is the least recognize curve. The most interesting finding was that there is a trend in misclassification. When misclassification takes place, forward curves are always misclassified as spiral curve and never as backward curve. Whereas spiral curves are always misclassify as either forward curve or backward curve. It seems possible that these results are due to the pressure profile of the curves. Backward curve has a high → low → high pressure profile and forward curve has a low → high → low pressure profile. Spiral curve has a complex pressure profile; the middle portion of the curve has a low → high → low pressure profile while the beginning and ending of the curve has no particular pressure trend. Among the three curves, backward curve has a unique pressure profile (high → low → high), which makes it distinct from forward and spiral curves and hence easily recognized. This factor may explain the relatively high recognition rate. Both forward and spiral curves have low → high → low pressure profile. The observed misclassification trend of the forward curve as spiral curve could be attributed to the low → high → low pressure profile they share. A possible explanation for the misclassification of spiral curves as either forward curve or backward curve may be the lack of consistency in pressure trend at the beginning and ending of the spiral curve. When high pressure is applied at the beginning and ending of the curve it is misclassified as a backward curve, whereas when low pressure is applied at the beginning and ending of the curve it is misclassified as a forward curve.

## 5 SIGNAL PROCESSING & SMOOTHING

### 5.1 Controlling Pressure

Like any other digital instruments, the pressure sensitive pen is very sensitive. It is responsive and able to capture the lightest pressure value above zero. Pressure value is captured from the movement the pen touches the sensing surface until it is completely lifted from the sensing surface. Such great sensitivity is a dilemma for cases where continuous pressure control is mapped to the control of a continuous parameter, depth distance in our case. Unwanted pressure input is captured when the pressure sensitive pen landed on the sensing surface before it reaches the desired starting pressure value. In this study, we refer to this as a landing effect. The action of lifting a pen, pencil or a brush away from a paper is so natural in real life, but not in the realm of the electronic medium. The action of lifting a pressure sensitive pen away from the sensing surface does not conform to human's perception of pressure and prior experience on pressure interaction. The pressure sensing and capturing by the pen based system does not stop at the last intended force applied. Pressure is still being captured throughout the action of lifting the pressure sensitive pen until the pressure value decreases to zero; when the pressure sensitive pen is completely lifted. As a result the action of lifting the pressure sensitive pen leaves a trail of sudden drop in pressure value, which we refer as landing effect in this study. This finding is consistent with those of Ramos et al. [8] and Mizobuchi et al. [7] who reported that the pen is very sensitive at low pressure level. The landing issue also accords with the observations by Ramos and Balakrishnan in [11]. Both landing and lifting effect described have produced significant distortion to the beginning and ending of all three curves; spiral curve, forward curve and backward curve

#### 5.1.1 Spiral Curve

An ideal spiral curve, in the context of this paper, consists of two aligned vertical edges and a slightly slanted oval or circle in the middle, as shown in Figure 1(a). The top view of a spiral takes the shape of the circle. The ideal pressure profile of a spiral is a smooth transition of pressure in a low → high → low manner, as illustrated in the dotted red line in Figure 2 (a). To create the aligned vertical edges of a spiral curve, constant low pressure is applied (depicted by the orange sphere) to ensure the two edges of the spiral is situated in the same depth position when the pressure profile is mapped to the depth distance. The circle shape in the middle of the spiral curve is created by increasing and decreasing the pressure level symmetrically during drawing. In reality, it is difficult to create the two vertical edges by maintaining the same pressure level while dragging a stylus. Unintended minor force area. For efficient pressure control, the decrement in pressure should be applied to the pressure signal at the intended area only, from finger tips tremor produces magnified pressure variation in pressure profile. The vertical edges created through the direct mapping of pressure data to depth distance do not have straight vertical profile. It is difficult to perform symmetric bi-directional pressure control to create a circle, thus the top view of the spiral is not in circular shape. The 90 degrees side view of the spiral has a dropping circle. The dropping circle is mainly caused by the nature of drawing the 2D circle of the spiral in x-y position. In addition, poor bi-directional pressure control of high → low pressure transition produces the issue of early crossing (too little pressure released, pressure applied is higher than required) and late crossing (too much pressure released, pressure applied is lower than required) in the spiral curve created. All the spiral curve pressure profiles captured from different users have the same pressure pattern or signature as illustrated in black solid line in Figure 2(a).

Table 2: Detail classification result for 50:35:3 architecture

| Curve Class | Total Test Performed | Classified Curve | | | Total Number of Correct Classification | Total Number of Misclassification | Correct Classification Rate |
|---|---|---|---|---|---|---|---|
| | | backward | forward | spiral | | | |
| **1HL 35 Neuron Classification Result** | | | | | **Error rate = 0.01063** | | |
| backward | 100 | 100 | 0 | 0 | 100 | 0 | 100% |
| forward | 100 | 0 | 98 | 2 | 98 | 2 | 98% |
| spiral | 100 | 1 | 4 | 95 | 95 | 5 | 95% |

### 5.1.2 Forward Curve

A forward curve is created by varying the pressure in low→high→low manner during sketching. Figure 1 (b) shows a schematic diagram of a forward curve. High pressure is the dominant pressure that controls the curvature of the 3D curve towards the user or camera. The forward curve pressure profiles captured from various users share the same pressure signature, as shown in black solid line in Figure 2(b). As can be seen in Figure 2(b), due to difficulty in pressure control, the application of high pressure not only increases the pressure of the intended area but span across the neighbouring pressure signal. To give better sense of pressure control, the increase in pressure should be applied to the pressure profile of the intended area only, as shown in the red dotted line graph in Figure 2 (b). In this study, we refer to this as localize pressure control at high pressure range.

### 5.1.3 Backward Curve

A backward curve is created by varying the pressure in high → low→high manner during sketching. Figure 1(c) shows a schematic diagram of a backward curve. Low pressure is the dominant pressure that controls the curvature of the 3D curve away from the user or camera. The backward curve pressure profiles captured from various users share the same pressure signature, as shown in black solid line in Figure 2(c). The application of low pressure not only decreases the pressure at the intended area, decrement in pressure signal span across a wide as shown in the dotted red line in Figure 2 (c). We refer to this as localize pressure control at low pressure range in this study.

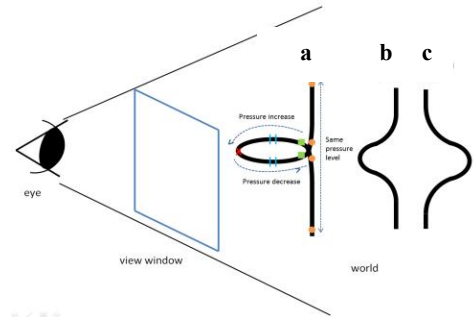

Figure 1: The ideal (a) spiral curve (b) forward curve (c) backward curve relative to user view point

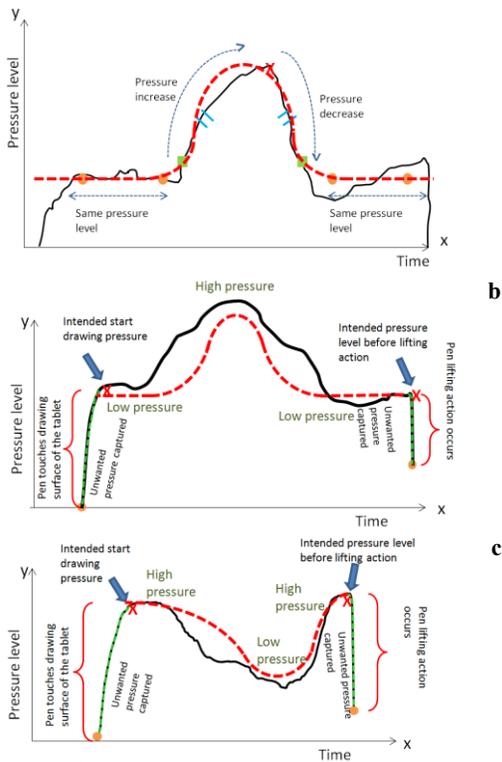

Figure 2: The ideal (red dotted line) and sketched (black solid line) pressure profile of a (a) spiral curve (b) forward curve (c) backward curve

## 5.2 Signal Processing

We experimented with the most promising signal processing techniques proposed in previous pressure based interaction studies [6, 8-14, 28] to observe the effect of the application of each technique on different curve categories. The signal processing techniques experimented include low pass filter, hysteresis filter, fisheye function, and sigmoid function. In the following graphs, the raw pressure profile is shown in blue line while the processed pressure profile is presented in red line.

### 5.2.1 Spiral Curve

- Low Pass Filter

In low pass filter, smoothing is achieved with the cost of phase shift. Therefore the low pass filter is only utilized for minor smoothing to avoid phase shift. The insufficiency of such minor smoothing is that the processed pressure signal still maintains the same pressure profile as the raw pressure signal.

- Hyteresis Filter

Our hysteresis implementation is based on the hysteresis filter proposed by Ramos et al. [11]. Good pressure control is observed from the use of hysteresis filter, it improves pressure transition without phase shifting. From the top view (see Figure 3(b)), the circular shape of the spiral has improved. Our result is consistent with previous study [11] which has reported that hysteresis filter works well in stabilizing pressure signal. Though hysteresis filter produces significant improvement in pressure control, the pressure stabilization and signal smoothing effect is achieved with large suppression on the pressure signal, as shown in Figure 3(a). The large suppression on pressure has greatly reduced the pressure difference between high and low pressure, resulted in a dropping and flat spiral curve, as shown in Figure 3(c). As a result, hysteresis filter is not suitable for spiral curve.

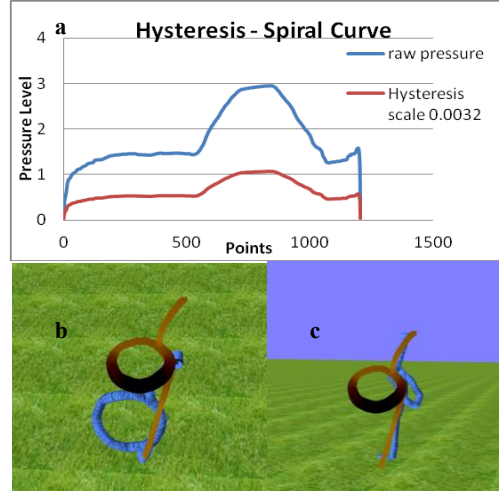

Figure 3: Spiral curve processed with hysteresis filter (a) pressure profile (b) top view (c) side view

- Fisheye Function

Our implementation is based on the fisheye function proposed in [10]. Both [10] & [28] reported good bi directional pressure control through the use of fisheye function proposed in [10]. The use of fisheye function in both studies is to discretize the pressure into smaller number of discrete pressure level for better target selection [10] and zooming purposes [28]. Other than discretizing the pressure signal, the fisheye function also scales the pressure signal relative to the maximum number of discrete pressure level set. Since our main goal of pressure signal processing is to produce smoother pressure transition, the discretization effect brought by fisheye function proposed in [10] is not suitable for continuous parameter control, as shown in Figure 4(a). The 3D spiral curve generated has been greatly distorted into discrete depth interval. The fisheye function is modified for continuous pressure control with the ability to cater for any number of pressure levels without exceeding the magnitude of the original pressure signal. Figure 4(b) shows the effect of applying the modified fisheye function on the pressure profile. The raw pressure signal is smoothed and suppressed. The moderate suppression in pressure produces better looking spiral without flattening the spiral as compared to hysteresis filter. From the 360 degrees views of the processed spiral curve, we can see that the circular shape of the spiral looks rounder and tighter; not as expand and out of shape as before processing. Both hysteresis and fisheye filter suppresses and smoothes the pressure without improving the pressure profile towards the ideal pressure profile of a spiral curve

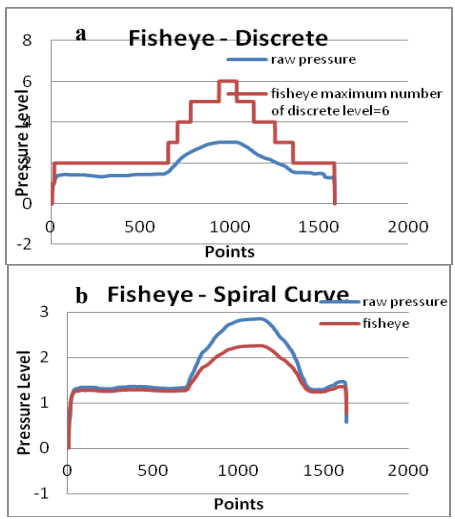

Figure 4: Spiral curve processed with fisheye function (a) discrete (b) modified

- Sigmoid Function

Sigmoid function has been investigated in few previous studies [9, 11, 14] to minimize the difference in pressure level perceived by human and sensed by digital instrument. Studies [9, 11, 14] concluded that sigmoid transfer function produces pretty good user feel for continuous pressure control, it is able to achieve the effect of pressure level change perceived by human when applying force to a pressure sensitive pen. In our implementation, the conventional sigmoid function was modified to include a contrast factor and threshold value. The beauty of the modified sigmoid function is that it is able to achieve 3 effects 1) localize pressure control at high pressure range 2) moderately suppress the high pressure signal 3) contrast enhancement at the two edges to make the pressure signal become almost a straight horizontal line. The major drawback of this function is that the pressure localization effect at the high pressure signal range changes the shape of the intended spiral curve. The processed pressure graph has a pointy peak, as shown in Figure 5. From the 360 degrees view of the 3D spiral curve generated, it is obvious that the localize pressure control at high pressure range has distorted the circular shape of the spiral. Reducing the contrast factor of the sigmoid function produces a smoother peak but with the cost of reducing the pressure difference between high and low pressure, resulted in flatten or dropping spiral curve.

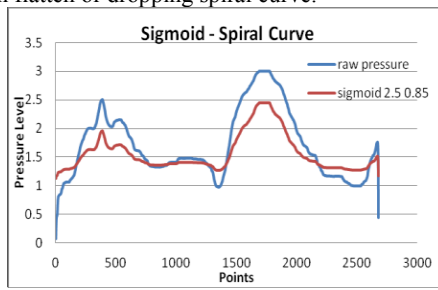

Figure 5: Spiral curve pressure profile processed with sigmoid function

- Spiral Processing

The existing signal processing techniques: low pass, hysteresis, fisheye, sigmoid are not sufficient to improve the pressure profile of a spiral curve. We designed a solution which is able to process the spiral curve pressure profile towards its ideal pressure profile. The designed solution comprises of 3 simple processes:

1. Process both horizontal edges of the spiral curve pressure profile towards an identified baseline; increase or decrease the pressure towards the baseline. The semicircular shape pressure profile of the spiral remains unprocessed.
2. Remove signal spike of the entire pressure profile using median function
3. Smooth the entire pressure profile using moving average

To determine the baseline, we analyzed various spiral curve pressure signal data captured from various users to identify the pressure signal trend in sketching a pressure based 3D spiral curve. Through our analysis, the desired pressures of both horizontal edges are usually situated at the median of the sorted spiral curve pressure profile. Therefore the median pressure value is used as the baseline for both horizontal edges. After obtaining the baseline value, we traced from both side of the peak to identify pressure values within the median pressure range. Contrast enhancement was performed on the identified pressure value, as shown in Figure 6. The formulated solution is able to perform what the previously described modified sigmoid function can do without changing the semicircular shape of the spiral pressure profile. A median filter was then applied to remove spike and followed by moving average function to smooth the pressure profile. The final processed pressure profile is close to the ideal pressure profile of a spiral curve. We also observed that by processing the pressure profile close to the ideal pressure profile, the vertical edges are aligned even when the pressure applied is not symmetrical at the crossing area, as shown in Figure 7(a) and 7(b). From the 2D visual feedback in Figure 7(a) and 7(b), we can clearly see that the color shades on the right and left of the circular shape sketched is not symmetrical. This indicates that the pressure applied during sketching is not symmetric, which will lead to early crossing in this case. With the designed solution, if the applied pressure is not symmetric at the crossing area, two possible scenarios will happen: the two vertical edges are attached to each other (too much pressure is released) or detached (too little pressure is released) from each other. With this we are able to ensure the two vertical edges are aligned and avoid the unfavorable early crossing and late crossing issues described above. The designed solution also solves the landing and lifting issue in spiral curve. The proposed solution gives an almost ideal looking pressure profile of a spiral curve, but the produced circular shape of the spiral is elongated. This is particularly obvious from the side view and top view, as shown in Figure 7(c) and 7(d). This effect is due to high pressure value difference between the high pressure range and low pressure range. To reduce the large pressure value differences, we applied fisheye function to moderately suppress the pressure to produce a tighter and rounder looking circle.

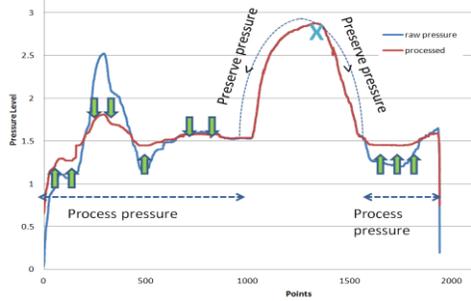

Figure 6: Perform contrast enhancement at both horizontal edges

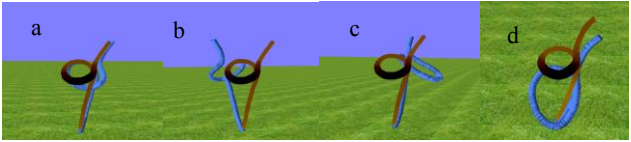

Figure 7: (a) Aligned vertical edges front view (b) aligned vertical edges back view (c) side view (d) top view

- The Combination of Signal Processing Techniques That Works Best for Spiral Curve

A combination of signal processing techniques is needed to improve pressure interaction. From the experimental studies above, we have formulated a combination of signal processing techniques to deal with the unique pressure interaction issues arise when sketching a spiral curve, as shown in Table 3. The combination of signal processing techniques is used in the identified order to maximize the strength of each signal processing techniques to improve pressure control. Low pass filter was first applied to perform minor smoothing on the raw pressure. This is then followed by our proposed solution. The beauty of this combination is that the major smoothing is concentrated at the two horizontal edges of the pressure profile, while the semicircular profile of the pressure curve is untouched. We can prevent unnecessary smoothing performed on the semicircular shape of the pressure profile; therefore maintain the intended drawn spiral shape. The median and moving average functions performed on the entire curve ensure the mountainous profile has a rounder peak and the horizontal edges are flatter. To improve bi-directional pressure control, the fisheye function was applied to moderately suppress the pressure signal. The proposed spiral signal processing was performed before the fisheye function. The outcome of doing this was that the 3D curve generated resembled what the user meant to draw and at the same time preserved the sketch details. If we suppress the pressure signal first, we will lose some distinct pressure profile or details of the drawing. Lastly, simple median and moving average function was performed on the y coordinate of the project 3D spiral to lift up the slightly dropping circular shape in the middle of the spiral curve. This combination of signal processing techniques works best in processing the spiral curve pressure profile into its ideal looking pressure profile, as demonstrated in Figure 8. Figure 9 shows multiple angle of the spiral curve created from the formulated combination of signal processing techniques. From the top view, we can see that the spiral takes the shape of the circle in the middle of the curve, as shown in Figure 10.

Table 3: Combination of signal processing techniques works best for spiral curve

| Signal Processing Techniques | Optimum parameter setting |
|---|---|
| Low pass filter | α = 0.075 |
| Spiral processing | NA |
| Fisheye function | l = 12, R = 600, r = 120, scale factor = 1/7, displacement = 0.65 |
| Median on Y coordinate of the projected 3D curve | NA |
| Moving average on Y coordinate of the projected 3D curve | NA |

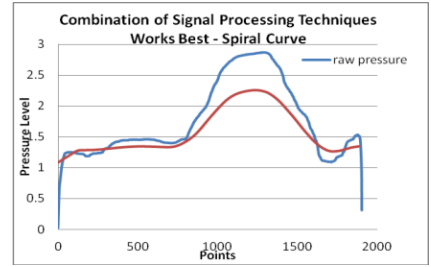

Figure 8: Spiral curve pressure profile processed with the combination of signal processing techniques works best

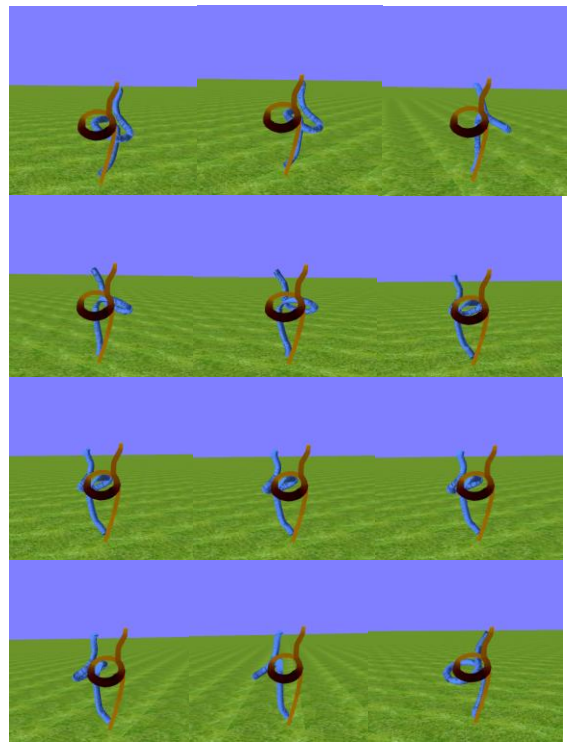

Figure 9: Spiral curve processed with the combination of signal processing techniques works best – multiple side view

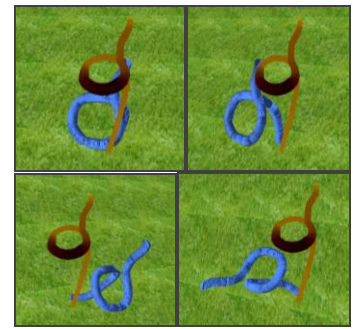

Figure 10: Spiral curve processed with the combination of signal processing techniques works best – multiple top view

### 5.2.2 Forward Curve

- Low Pass Filter

Similar to the spiral curve, low pass filter was applied to perform minor smoothing without introducing phase shift. From the experimental result, we discovered that the optimum parameter setting for spiral curve is not applicable to forward curve. There is an observable phase shift when the optimum parameter setting for spiral curve was used.

- Sigmoid Function

The same sigmoid function discussed earlier was used. We successfully demonstrated the use of our modified sigmoid function for localize pressure control at high pressure range with moderate pressure suppression, as shown in Figure 11. The application of high pressure only increased the pressure at the intended area. It suppressed the pressure signal at low pressure range. Through the processed pressure profiles and the 3D forward curves created, we also observed that the landing and lifting issues are solved.

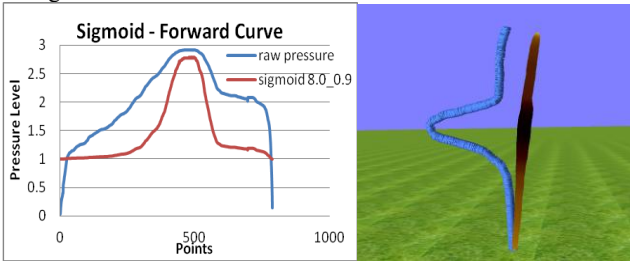

Figure 11: Forward curve processed with sigmoid function

- The Combination That Works Best for Forward Curve

From the experimental results obtained from spiral curves. We found that in order to improve pressure control, the raw pressure signal needs to be suppressed. The two signal processing techniques which can suppress the pressure signal without changing the shape of the pressure profile are hysteresis function and fisheye function. In finding the combination of signal processing techniques that works best for forward curve, we compared two combinations of signal processing techniques; each having either hysteresis or fisheye function as pressure suppression function. Both combinations of signal processing techniques are able to improve pressure control in sketching a forward curve, as shown in Figure 12. Fisheye function provides good bi directional pressure control, this observation is consistent with the research findings in both [10] & [28]. The use of hysteresis filter reduces the sensitivity in pressure control, in another words it can stabilize the pressure better than fisheye function. However the forward curve created by hysteresis function is stiff and flatter (see Figure 12(c)) as compared to fisheye (Figure 12(b)).

a

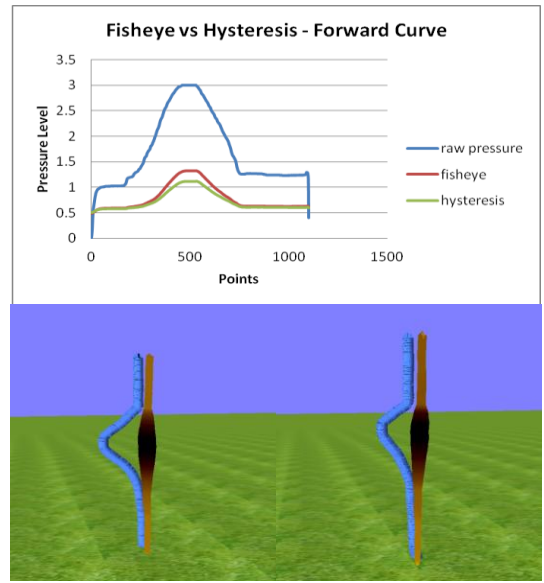

b             c

Figure 12: Forward curve processed with fisheye and hysteresis function (a) processed pressure profile: red-fisheye, green-hysteresis (b) fisheye (c) hysteresis

The combination of signal processing techniques and their parameter setting that works best for forward curve is listed in Table 4. First low pass filter was applied to the raw pressure signal to perform minor smoothing without phase shifting. Subsequently sigmoid function was applied to achieve localize pressure control at high pressure range before pressure suppression. Through our experimental studies, better pressure control was observed when the pressure value was localized at high pressure range before suppression. Fisheye function suppressed the pressure signal and at the same time improved bi-directional pressure control. Smoother pressure transition was observed with the use of fisheye function. Hysteresis function stabilized the pressure signal, but too much suppression caused the application of pressure did not conform to user's perception of pressure. Users applied big pressure differences during bi-directional pressure control to create a big curve, but it was not portrayed in the 3D curve created due to great pressure suppression, which reduced the depth distance and therefore flatten the curve. Conversely, fisheye provided good bi directional pressure control. It was more responsive and the increment and decrement in depth distance was what the user would have expected. We prefer the responsive effect provided by the fisheye function and the rounded curvature it can produce.

Table 4: Combination of signal processing techniques works best for forward curve

| Signal Processing Techniques | Optimum parameter setting |
|---|---|
| Low pass filter | α = 0.1 |
| Sigmoid function | contrast factor = 2.5, threshold = 0.85 |
| Fisheye function | l = 10, R = 600, r = 120, scale factor = 1/6, displacement = 0 |

### 5.2.3 Backward Curve

Low Pass Filter

We used the same experiment setting for forward curve to find the optimum parameter setting for backward curve. The backward curve and forward curve share the same optimum parameter setting, where minor smoothing is achieved without observable phase shift.

- Sigmoid Function

Previously we successfully demonstrated the use of sigmoid function for localize pressure control at high pressure range. In processing the forward curve, pressure localization at high pressure range was achieved by suppressing the low pressure signals. Pressure variation at low pressure range was flattened out, while the intended dominant high pressure signals are maintained. The sigmoid configuration used for forward curve was not applicable for backward curve. For forward curve, we tried to suppress the low pressure signals as much as possible, while in backward curve, the low pressure signals are the dominant pressure, we need to preserve the low pressure signals range observed in the raw pressure profile. As shown in Figure 13, we can see that when the same sigmoid configuration as forward curve was used, low pressure signal range and 3D backward curve generated was widened instead of localize. Therefore the sigmoid configuration used for forward curve is not applicable to backward curve.

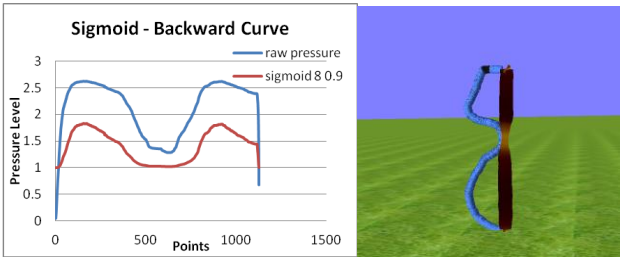

Figure 13: Backward curve processed with the optimum sigmoid configuration for forward curve

Through the experimental observations we found that when low contrast and low threshold value was used, we are able to minimize the pressure variation at high pressure range and suppress the pressure signal difference, as shown in Figure 14. Previously we have demonstrated that the use of the sigmoid function in forward curve to overcome the landing and lifting issues. However, in backward curve, the application of sigmoid function does not fully solve the landing and lifting issues, extended 'head' and 'tail' is observed, as shown in Figure 14.

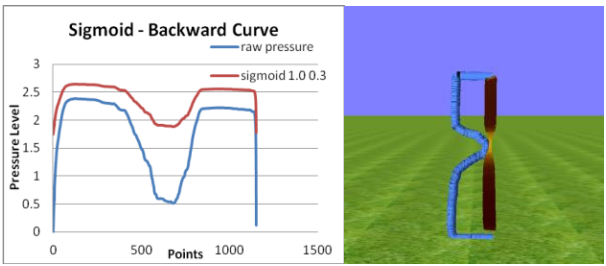

Figure 14: Backward curve processed with sigmoid function - low contrast and low threshold configuration

- Overcoming Landing & Lifting Issues

To overcome the landing and lifting issues, we studied the trend of the backward curve pressure profiles captured from various users. We tried to identify as accurately as we can the start drawing pressure to mitigate the landing effect. We designed an algorithm to trace from the first pressure value until the pressure stops its increasing trend. The first pressure value encountered after the increasing trend stops is referred as the start drawing pressure. We then replaced the increasing trend with the start drawing pressure. To mitigate the landing effect, we tried to identify as accurately as we can the last intended pressure before the pen lifting action occurs. From our observation, when the pen lifting action happened the pressure value from that point onwards followed a monotonically decreasing trend. Our algorithm traced the pressure value backwards until the pressure value before the decreasing trend started was found, we refer it as the stop drawing pressure. The pressure values after the stop drawing pressure point onwards were replaced with the stop drawing pressure.

- The Combination That Works Best for Backward Curve

The combination of signal processing techniques that works best in dealing with the pressure interaction issues that arise in sketching a backward curve is presented in Table 5. First, we mitigated the landing and lifting issues by reassigning the pressure values. Subsequently low pass filter was applied to perform minor smoothing without introducing phase shift. Followed by the application of sigmoid function to localize pressure control at low pressure range and suppress the pressure moderately. Similar experiment setting as forward curve was used in determining whether to use fisheye or hysteresis function for pressure suppression. Similar experimental result was obtained. Therefore fisheye function was used to suppress pressure and improve bi-directional pressure control for smoother pressure transition. However, different fisheye configuration was used, as shown in Table 5. Using the below combination of signal processing techniques, we observed good mapping of pressure to depth distance from the processed pressure profile (Figure 15(a)) and their corresponding 3D backward curve produced, as shown in the screenshots captured from various angle in Figure 15(b)

Table 5: Combination of signal processing techniques works best for backward curve

| Signal Processing Techniques | Optimum parameter setting |
|---|---|
| Reassign pressure value | NA |
| Low pass filter | $\alpha = 0.1$ |
| Sigmoid function | contrast factor = 1.0, threshold = 0.3 |
| Fisheye function | l = 10, R = 600, r = 120, scale factor = 1/5, displacement = 0 |

## 5.3 Smoothing

After obtaining a positive result that different type of curve needs different combination of signal processing techniques, out of our interest, we experimented few smoothing techniques to observe if different smoothing technique is needed to smooth each curve type. The smoothing techniques experimented include catmull rom spline, hermite, chaikin with filter width 4 and filter width 8, quadratic bezier, cubic bezier and b-spline. Chaikin with filter width 4 and 8, and b-spline work best in producing smooth and beautiful spiral curve while catmull rom spline generates good result for both forward curve and backward curve. We present the result of processing the spiral curve, forward curve and backward curve with its best combination of signal processing and smoothing techniques in Figure 16 to Figure 18 respectively.

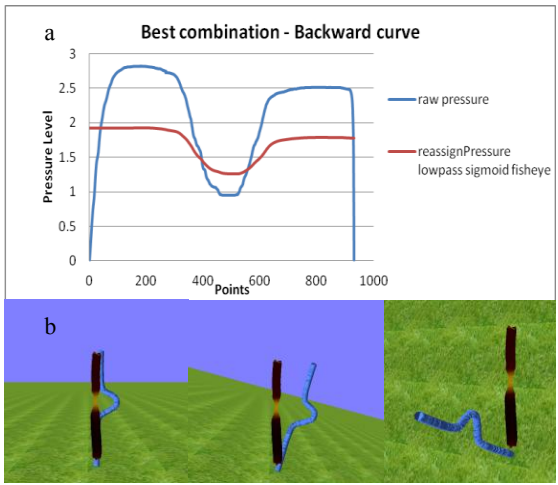

Figure 15: Backward curve processed with the best combination of signal processing techniques (a) processed pressure profile (b) processed backward curve - multiple angle

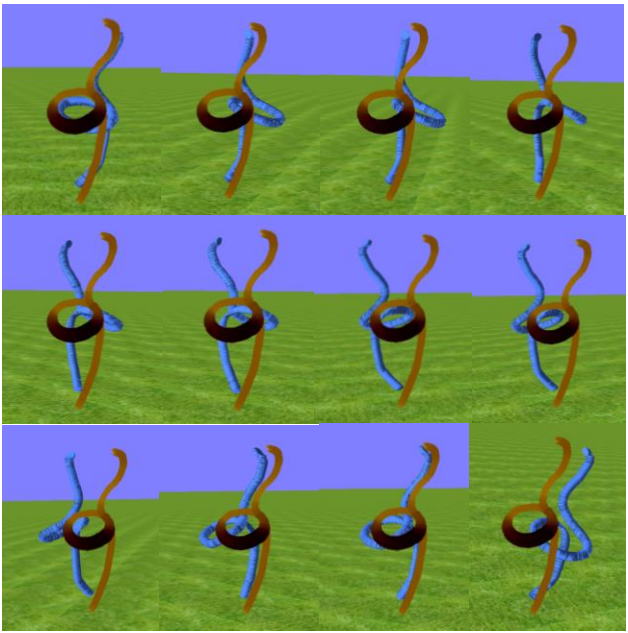

Figure 16: Spiral curve processed with the best combination of signal processing techniques and b-spline smoothing technique

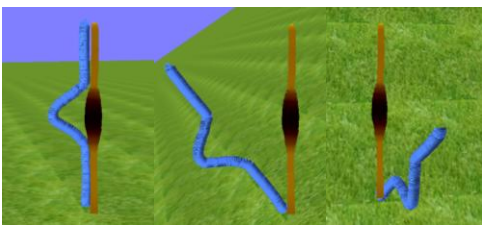

Figure 17: Forward curve processed with the best combination of signal processing techniques and catmull rom spline smoothing technique

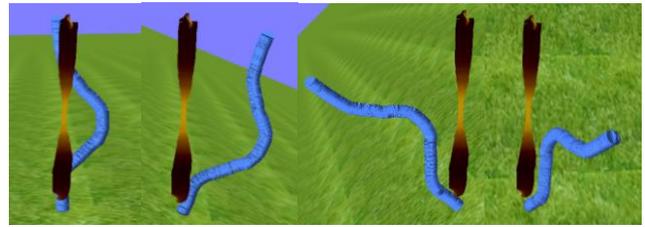

Figure 18: Backward curve processed with the best combination of signal processing techniques and catmull rom spline smoothing technique

## 6 CONCLUSION

We have presented in this paper a framework for 3D curve drawing that exploit curve pressure profiles. The framework is, in principle, simple and is in fact the underlying principle of sketch-based modelling system (such as [29]): *draw-recognize-process*. The user draws, and the system recognizes the drawing and process it accordingly. Our contribution is in developing the framework for a particular drawing task that we have assumed to be 'unsolved': drawing 3D curves. The contribution is significant especially when taking into consideration the difficulty of projecting 2D curves to 3D based on its pressure profiles.

While the framework has been developed for only a limited set of curve types (forward, backward, and spiral with a single loop), we believe that it is extendable and applicable to any curve type. In fact, the implementation of this extension is a work-in-progress.

The work bears implication for future sketch-based interface. Incorporating in such interface the ability to recognize and to process input sketch, especially curves, according to the recognition result, should lead to smoother user experiences and its wider acceptance.

### REFERENCES.

[1] Bae, S.-H., R. Balakrishnan, and K. Singh. ILoveSketch: as-natural-as-possible sketching system for creating 3d curve models. in Proceedings of the 21st annual ACM symposium on User interface software and technology. 2008. ACM.

[2] Cohen, J.M., et al. An interface for sketching 3D curves. in Proceedings of the 1999 symposium on Interactive 3D graphics. 1999. ACM.

[3] De Amicis, R., et al., The eraser pen: a new interaction paradigm for curve sketching in 3D. Dubrovnik, 2002.

[4] Grossman, T., et al. Creating principal 3D curves with digital tape drawing. in Proceedings of the SIGCHI Conference on Human Factors in Computing Systems. 2002. ACM.

[5] Schmidt, R., et al. On expert performance in 3D curve-drawing tasks. in Proceedings of the 6th Eurographics Symposium on Sketch-Based Interfaces and Modeling. 2009. ACM.

[6] Cechanowicz, J., P. Irani, and S. Subramanian. Augmenting the mouse with pressure sensitive input. in Proceedings of the SIGCHI conference on Human factors in computing systems. 2007. ACM.

[7] Mizobuchi, S., et al. Making an impression: force-controlled pen input for handheld devices. in CHI'05 extended abstracts on Human factors in computing systems. 2005. ACM.

[8] Ramos, G., M. Boulos, and R. Balakrishnan. Pressure widgets. in Proceedings of the SIGCHI conference on Human factors in computing systems. 2004. ACM.

[9] Ren, X., et al., The adaptive hybrid cursor: A pressure-based target selection technique for pen-based user interfaces, in Human-Computer Interaction–INTERACT 2007. 2007, Springer. p. 310-323.

[10] Shi, K., et al. PressureFish: a method to improve control of discrete pressure-based input. in Proceedings of the SIGCHI Conference on Human Factors in Computing Systems. 2008. ACM.

[11] Ramos, G. and R. Balakrishnan. Zliding: fluid zooming and sliding for high precision parameter manipulation. in Proceedings of the 18th annual ACM symposium on User interface software and technology. 2005. ACM.

[12] Ramos, G.A. and R. Balakrishnan. Pressure marks. in Proceedings of the SIGCHI conference on Human factors in computing systems. 2007. ACM.

[13] Shi, K., S. Subramanian, and P. Irani, PressureMove: Pressure Input with Mouse Movement, in Human-Computer Interaction–INTERACT 2009. 2009, Springer. p. 25-39.

[14] Barrett, R.C., R.S. Olyha Jr, and J.D. Rutledge, Graphical user interface cursor positioning device having a negative inertia transfer function, 1996, Google Patents.

[15] Ramos, G.A., *Pressure-sensitive pen interactions*, 2008, University of Toronto.

[16] Subasi, A. and E. Erçelebi, *Classification of EEG signals using neural network and logistic regression.* Computer Methods and Programs in Biomedicine, 2005. **78**(2): p. 87-99.

[17] Rajendra Acharya, U., et al., Classification of heart rate data using artificial neural network and fuzzy equivalence relation. Pattern Recognition, 2003. **36**(1): p. 61-68.

[18] Ceylan, M., et al., Classification of carotid artery Doppler signals in the early phase of atherosclerosis using complex-valued artificial neural network. Computers in Biology and Medicine, 2007. **37**(1): p. 28-36.

[19] Ceylan, R., et al., *Fuzzy clustering complex-valued neural network to diagnose cirrhosis disease.* Expert Systems with Applications, 2011. **38**(8): p. 9744-9751.

[20] Engin, M., *ECG beat classification using neuro-fuzzy network.* Pattern Recognition Letters, 2004. **25**(15): p. 1715-1722.

[21] Osowski, S. and T.H. Linh, *ECG beat recognition using fuzzy hybrid neural network.* Biomedical Engineering, IEEE Transactions on, 2001. **48**(11): p. 1265-1271.

[22] Özbay, Y., R. Ceylan, and B. Karlik, *A fuzzy clustering neural network architecture for classification of ECG arrhythmias.* Computers in Biology and Medicine, 2006. **36**(4): p. 376-388.

[23] Özbay, Y. and G. Tezel, A new method for classification of ECG arrhythmias using neural network with adaptive activation function. Digital Signal Processing, 2010. **20**(4): p. 1040-1049.

[24] Sternickel, K., *Automatic pattern recognition in ECG time series.* Computer methods and programs in biomedicine, 2002. **68**(2): p. 109-115.

[25] Tezel, G. and Y. Özbay. A new neural network with adaptive activation function for classification of ECG arrhythmias. in Knowledge-Based Intelligent Information and Engineering Systems. 2007. Springer.

[26] Kotyrba, M., et al., ELLIOTT WAVES RECOGNITION VIA NEURAL NETWORKS.

[27] Lai, C.-Y. and N. Zakaria. *Pressure-based 3D curve drawing*. in *Smart Graphics*. 2010. Springer.

[28] Mandalapu, D. and S. Subramanian. Exploring pressure as an alternative to multi-touch based interaction. in Proceedings of the 3rd International Conference on Human Computer Interaction. 2011. ACM.

[29] Igarashi, T., S. Matsuoka, and H. Tanaka. Teddy: a sketching interface for 3D freeform design. in ACM SIGGRAPH 2007 courses. 2007. ACM.